# On Neutrinos and Fermionic Mass Patterns

Paul M. Fishbane
*Physics Dept. and Institute for Nuclear and Particle Physics,*
*Univ. of Virginia, Charlottesville, VA 22903*

Peter Kaus
*Physics Dept., Univ. of California, Riverside, CA 92521*

Abstract

Recent data on neutrino mass differences is consistent with a hierarchical neutrino mass structure strikingly similar to what is observed for the other fermionic masses.

Recent experiments [1] have provided quantitative results on differences of the neutrino masses squared,

$$\Delta_{\tau\mu} = m_{\nu\tau}^2 - m_{\nu\mu}^2 \text{ and } \Delta_{\mu e} = m_{\nu\mu}^2 - m_{\nu e}^2. \tag{1a}$$

These quantities are given numerically by

$$\Delta_{\tau\mu} = (2 \pm 1) \times 10^{-3} \text{ eV}^2 \text{ and } \Delta_{\mu e} = (6 \pm 1.5) \times 10^{-6} \text{ eV}^2. \tag{1b}$$

This data appears to be consistent in the context of three neutrinos with other experimental information available [2, 3], if not with all of it [4; to accomodate all the data, one requires additional neutrino(s)—see ref. 5].

In thinking about how to use this data to best advantage, we can start with the conservative assumption that neutrinos come in three families, with no additional species, sterile or otherwise. Of course the two pieces of data are inadequate to give the three neutrino masses. We can get further guidance on how to proceed by looking at the other fermion sectors, both charged leptons and quarks. There is one obvious common feature of the different sectors, and that is that the masses have a hierarchical structure; that is, $m_1 \ll m_2 \ll m_3$, where $m_1$ refers respectively to the $d$-quark, $u$-quark, and electron in the down-quark, up-quark, and charged lepton sectors. When the quark masses are run within the minimal supersymmetric model to unification scales [6] the mass ratios are:

down-quark sector:          $m_s/m_b = 0.034$, $m_d/m_s = 0.049$
up-quark sector:             $m_c/m_t = 0.0035$, $m_u/m_c = 0.0038$       (2a)
charged lepton sector:      $m_\mu/m_\tau = 0.059$, $m_e/m_\mu = 0.0048$

Not only do the charged fermion sectors have a hierarchical mass structure, but it is also true that this structure can be expressed as powers of a single parameter $\lambda$, the sine of the Cabibbo angle, or the Wolfenstein parameter[7], roughly 0.22. Powers of this quantity appear throughout the CKM matrix with coefficients of order 1, and the same is true for the mass ratios, which can be written up to coefficients of order 1 as

down-quark sector:          $m_b:m_s:m_d = 1:\lambda^2:\lambda^4$
up-quark sector:             $m_t:m_c:m_u = 1:\lambda^4:\lambda^8$               (2b)
charged lepton sector:      $m_\tau:m_\mu:m_e = 1:\lambda^2:\lambda^6$

This fact can be quantified in several ways. For example, we can define $\lambda$ according to these ratios and see to what extent the same values of $\lambda$ appear. In this way we find

$$\lambda_{db} \equiv \left(\frac{m_d}{m_b}\right)^{1/4} = 0.202; \quad \lambda_{sb} \equiv \left(\frac{m_s}{m_b}\right)^{1/2} = 0.184; \quad \lambda_{ds} \equiv \left(\frac{m_d}{m_s}\right)^{1/2} = 0.220$$

$$\lambda_{ut} \equiv \left(\frac{m_u}{m_t}\right)^{1/8} = 0.245; \quad \lambda_{ct} \equiv \left(\frac{m_c}{m_t}\right)^{1/4} = 0.243; \quad \lambda_{uc} \equiv \left(\frac{m_u}{m_c}\right)^{1/4} = 0.248 \quad (3)$$

$$\lambda_{e\tau} \equiv \left(\frac{m_e}{m_\tau}\right)^{1/6} = 0.257; \quad \lambda_{\mu\tau} \equiv \left(\frac{m_\mu}{m_\tau}\right)^{1/2} = 0.244; \quad \lambda_{e\mu} \equiv \left(\frac{m_e}{m_\mu}\right)^{1/4} = 0.264.$$

The consistent presence of powers of $\lambda$ is striking. Even if the particular power of $\lambda$ is different in each case, those powers appear in simply related form. This of course may only be numerical coincidence, a remark that would be true for the CKM matrix as well. If it is, we shall find below that the coincidence has been extended to the neutrino sector.

As for absolute values, the charged lepton masses do not appreciably run, so that we can use the value $m_\tau = 1.78$ GeV. The very stable grand unification relation $m_b = m_\tau$ gives us a convenient route to the down quark sector. However, the absolute scale for the up sector cannot be known reliably, because the top quark mass is at the infrared fixed point of the theory and is therefore insensitive to an initial value at ultraviolet scales; when below we have occasion to refer to the top quark mass we use 300 GeV.

We most easily check to what extent the neutrino masses can fit into this picture by defining the hierarchy parameter $R$ as follows:

$$m_{\nu e}/m_{\nu \mu} = R \, m_{\nu \mu}/m_{\nu \tau} \quad (4)$$

Of course, $R$ only parameterizes solutions to Eq. (1a), but it helps to define what we mean by a "hierarchical solution." A solution to Eqs. (1a) and (4) is hierarchical when $R$ is very close to 1 or when it is close to a power of a (small) mass ratio, here $m_{\nu\mu}/m_{\nu\tau}$. We see from Eq. (2) that the analog to $R$ for both the down and up quarks is very close to unity (1.43 and 1.09, respectively). The charged leptons are associated with an analog to $R$ of $O(\lambda^2) = O(m_\mu/m_\tau)$.

Is one of the (quadratic) solutions for the square of the masses hierarchical? The answer is yes. One branch not only gives a small value of $m_{\nu\mu}/m_{\nu\tau}$ but also permits both $R = 1$ and $R = m_{\nu\mu}/m_{\nu\tau}$ as well as $R = 0$ (leading to a massless electron neutrino). Moreover, for this branch the ratio $m_{\nu\mu}/m_{\nu\tau}$ is nearly independent of $R$. This is the branch that interests us. (See Ref. [8] for a the second, nonhierarchical, solution.) The solution on the branch of interest is

$$m_{\nu_\mu} = \frac{\left(\sqrt{-4R^2\Delta_{\tau\mu}\Delta_{\mu e} + (\Delta_{\tau\mu} + \Delta_{\mu e})^2} - \Delta_{\tau\mu} + \Delta_{\mu e}\right)^{1/2}}{\sqrt{2(1-R^2)}} \quad (5a)$$

$$m_{\nu_\tau} = \sqrt{\Delta_{\tau\mu} + \mu_{\nu_\mu}^2} \quad (5b)$$

$$m_{\nu_e} = \frac{R\mu_{\nu_\mu}^2}{\sqrt{\Delta_{\tau\mu} + \mu_{\nu_\mu}^2}} \quad (5c)$$

with an appropriate continuation for $R > 1$.

For the whole range of $R$ that is allowed for this branch $\left(0 < R < \frac{1}{2}\frac{\Delta_{\tau\mu} + \Delta_{\mu e}}{\sqrt{\Delta_{\tau\mu}\Delta_{\mu e}}}\right)$, the ratio $m_{\nu\mu}/m_{\nu\tau}$ is indeed insensitive to $R$. This is evident in the solution at some special values. The case $R = 1$, which the quark sector suggests, gives

$$\mu_{\nu_\tau} = \frac{\Delta_{\tau\mu}}{\sqrt{\Delta_{\tau\mu} - \Delta_{\mu e}}} \cong \sqrt{\Delta_{\tau\mu}};$$

$$\mu_{\nu_\mu} = \frac{\sqrt{\Delta_{\tau\mu}\Delta_{\mu e}}}{\sqrt{\Delta_{\tau\mu} - \Delta_{\mu e}}} \cong \sqrt{\Delta_{\mu e}}; \quad (6)$$

$$\mu_{\nu_e} = \frac{\Delta_{\mu e}}{\sqrt{\Delta_{\tau\mu} - \Delta_{\mu e}}} \cong \frac{\Delta_{\mu e}}{\sqrt{\Delta_{\tau\mu}}}$$

The hierarchical nature of these forms is clear when we recall that $\Delta_{\tau\mu} \gg \Delta_{\mu e}$. We have indeed used this to get the approximate forms. The other interesting special case is $R = m_{\nu\mu}/m_{\nu\tau}$, (i.e., $m_{\nu e}/m_{\nu\mu} = (m_{\nu\mu}/m_{\nu\tau})^2$, as in the charged lepton sector). This case gives the approximate result

$$m_{\nu\tau} \cong (\Delta_{\tau\mu})^{1/2}; \; m_{\nu\mu} \cong (\Delta_{\mu e})^{1/2}; \; m_{\nu e} \cong (\Delta_{\mu e})^{3/2}/\Delta_{\tau\mu}. \quad (7)$$

Here we have used $\Delta_{\tau\mu} \gg \Delta_{\mu e}$ to approximate an otherwise opaque result. These $\nu_\tau$ and $\nu_\mu$ masses are roughly the same as the $R = 1$ case, Eq. (6).

Since in this solution the $\nu_\tau$ and $\nu_\mu$ masses are given nearly independently of $R$ by the data of Ref. [1], so is their mass ratio,

$$m_{\nu\mu}/m_{\nu\tau} \cong (\Delta_{\mu e})^{1/2}/(\Delta_{\tau\mu})^{1/2} = 0.055 \pm 0.014. \quad (8)$$

Here we have taken the numerical values of Eq. (1b) with errors treated by quadrature.

*This numerical value is very close to the ratio of the two heaviest masses in the down and charged lepton sectors.* If, as those sectors suggest, we define a quantity $\lambda_\nu$ through $m_{\nu\mu}/m_{\nu\tau} = \lambda_\nu^2$, then

$$\lambda_\nu = 0.234 \pm 0.029. \tag{9}$$

This value spans the $\lambda$-values defined by the other three sectors, as Fig. 1 illustrates.

We conclude our discussion with some remarks about the masses themselves. We have stated that the hierarchical solution determines the masses of the $\tau$- and $\mu$-neutrinos themselves; to a good approximation these are given by Eq. (6), and imply the numerical values

$$m_{\nu\tau} = (4.5 \pm 1.1) \times 10^{-2} \text{ eV and } m_{\nu\mu} = (2.5 \pm 0.3) \times 10^{-3} \text{ eV}. \tag{10}$$

The $m_{\nu\tau}$ result can be compared to two recent $O(10^{-2}$ eV) predictions [9], both of which extract the scale for the neutrino masses from the unification scale for a supersymmetric grand unified theory.

The value of $m_{\nu e}$ depends of course on $R$. If the neutrino sector follows the down-quark pattern, then its mass is $\lambda_\nu^2$ times $m_{\nu\mu}$. If it follows the charged lepton sector then its mass is $\lambda_\nu^4$ times $m_{\nu\mu}$. We can refer to these possibilities as case a and case b respectively. In Fig. 2 we plot the logarithms of these masses as a function of family, together with the same quantity for the other three sectors. This graph reveals just how different the neutrino sector must be from the other three sectors in absolute value and how similar all the sectors could be in slopes. The difference may be easier to understand than the similarity. Even within a purely standard model view of neutrinos—three sets of right-handed singlets and left-handed doublets—they differ from the other fermions in that they have no electrical charge and can get mass either by a Dirac term alone or by Dirac and Majorana terms that combine through a seesaw mechanism. The appearance of a common value of $\lambda$ in this context appears to us to be remarkable.

We refer the reader elsewhere [8; see also 10] for a discussion of the leptonic analog to the CKM matrix, as well as of models that could shed light on the results described here.

## Acknowledgments

We thank P. Ramond for many useful conversations, Darrel Smith for much assistance and the Aspen Center for Physics for its hospitality. PMF is supported in part by the U.S. Department of Energy under grant number DE-AS05-89ER40518.

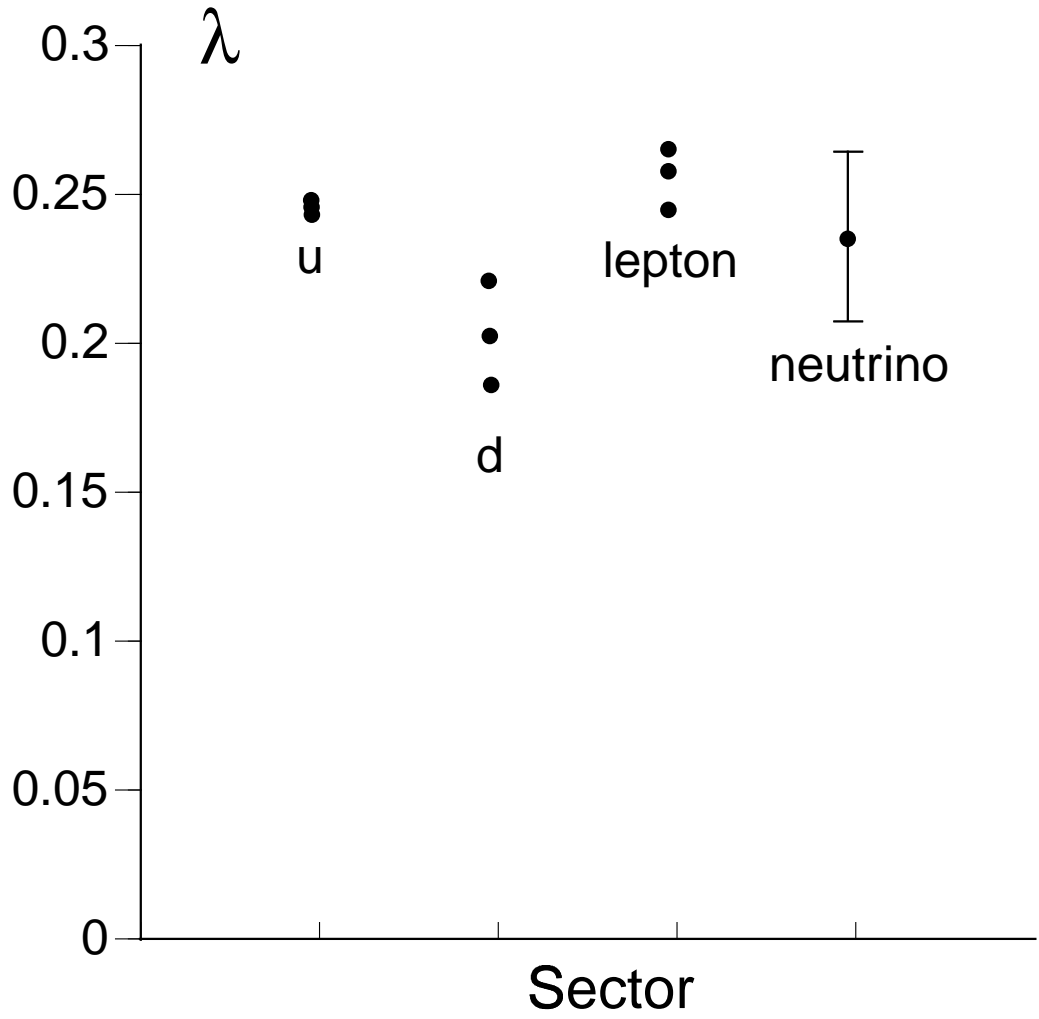

**Figure 1**

The values of $\lambda$ extracted from mass ratios according to Eqs. (3) and (9).

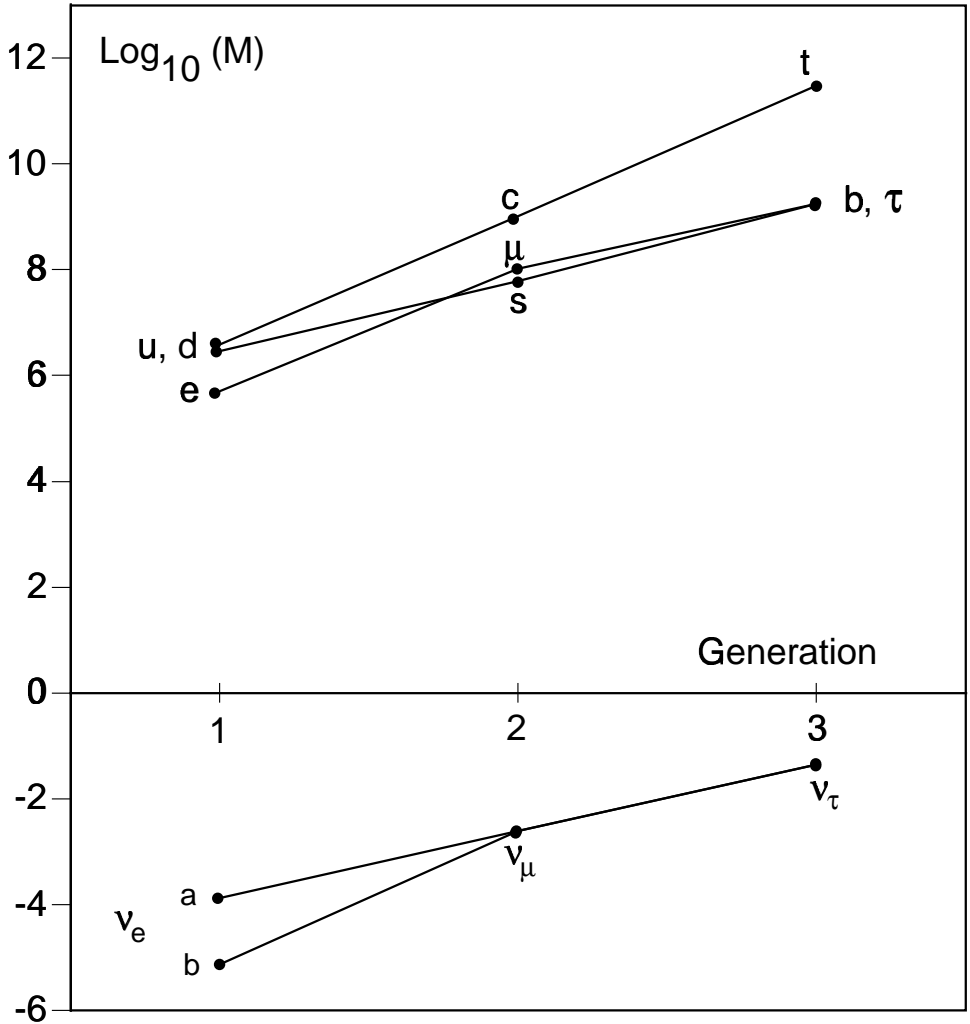

**Figure 2**

The logarithms of the fermion masses. Solid lines connect these masses within a given sector, enabling a visual comparison of the sect0ors. For the neutrinos, the labels a and b refer to a choice of $R = 1$ and $R = O(\lambda^2)$, according to the text discussion.